\documentclass[aps,prd,reprint,preprintnumbers,nofootinbib]{revtex4-1}

\usepackage{amsmath}
\usepackage{amsfonts} 
\usepackage{amssymb}
\usepackage{graphicx}
\usepackage{hyperref}
\usepackage{tensor}
\usepackage{siunitx}
\usepackage{float}
\usepackage{bm}
\usepackage{mathtools}
\DeclarePairedDelimiter{\floor}{\lfloor}{\rfloor}
\newcommand{\be}{\begin{equation}}
\newcommand{\ee}{\end{equation}}

\begin{document}

\title{\Large Non-perturbative topological string theory on compact \\ Calabi-Yau manifolds from M-theory}

\author{\large Jarod Hattab}
\author{\large Eran Palti}

\affiliation{\vspace{0.4cm}
Department of Physics, Ben-Gurion University of the Negev, Beer-Sheva 84105, Israel}

\begin{abstract}
We show that the full non-perturbative topological string free energy, in the holomorphic limit, follows simply from a target space integrating out calculation of M2 states. Qualitatively, this is the same as the calculation performed by Gopakumar and Vafa, but we find that the final expression must be modified due to a subtlety with poles induced by non-perturbative physics. Accounting for this modification leads to a Gopakumar-Vafa-like formula, which we propose as the exact formulation of the integrating out procedure. Evaluating the formula necessarily requires a contour integral in a complexified Schwinger proper time parameter. We show that this evaluation yields the full non-perturbative topological string free energy, and can be applied to a compact, or non-compact, Calabi-Yau threefold. The explicit formula presented holds as long as the two-cycles wrapped by the branes are rigid and smooth, but the methodology can be used to study also more general Calabi-Yau geometries. 
\vspace{1cm}
\end{abstract}

\maketitle

\section{Introduction}
\label{sec:int}

Topological string theory was first defined as a perturbative theory \cite{Witten:1988xj,Vafa:1990mu,Witten:1991zz} (see \cite{Neitzke:2004ni,Vonk:2005yv,Marino:2005sj} for reviews). The A-model topological string free energy $F^p$ is defined perturbatively through a genus expansion
\begin{eqnarray}
    F^p = \sum_{g = 0}^{\infty}F_g\lambda^{2g-2}\;,
\end{eqnarray}
as an asymptotic series in the topological string coupling $\lambda$. The genus $g$ prepotential $F_g$ roughly counts the number of holomorphic maps from genus $g$ Riemann surfaces to the Calabi-Yau $Y$ weighted by powers of $e^{-\beta\cdot t}$ where $\beta\cdot t = \sum_i \beta_it^i$ is the area of the cycle in $Y$ under the given holomorphic map. The $t^i$ parameterise a complexified Kahler form in a basis of $H_2(Y,\mathbb{Z})$ and $\beta_i \in \mathbb{N}$ give the class of the image in this basis. 

In \cite{Gopakumar:1998ii,Gopakumar:1998jq} Gopakumar and Vafa showed using relations between type IIA string theory compactified on $Y$ and the A-model on $Y$ that $F^p$ encodes the BPS structure of wrapped M2-branes in an M-theory compactification on $Y$. 
More precisely, certain F-terms $I_g$ in the effective action for type IIA on $Y$ were known to be computed by the genus $g$ topological string partition functions \cite{Antoniadis:1993ze}
\begin{eqnarray}
    I_g = -i\int \text{d}^4x\text{d}^4\theta F_{g}(\mathcal{X}^{I})\mathcal{W}^{2g}\;,
\end{eqnarray}
where $\mathcal{X}^I$ and $\mathcal{W}$ are superfields whose bottom components are scalars $t^i$ and the (anti-selfdual) graviphoton field strength $W^-$. The idea was to reinterpret $\sum_{g = 0}^{\infty}I_g$ as the asymptotic expansion of the superspace effective action obtained from integrating out, in Schwinger's formalism, specific BPS particle excitations of wrapped M2 branes in a constant anti-selfdual graviphoton field on $\mathbb{R}^4$. The integrating out calculation was argued to yield an expression for the exponential parts of the perturbative (in $\lambda$) topological string free energy \cite{Gopakumar:1998jq}
\begin{eqnarray}
\label{FP}
    F^p = \sum_{{\bf \beta},r\geq 0,n\in \mathbb{Z}}\alpha_r^{{\bf \beta}}\int_{\epsilon}^{\infty}\frac{\text{d}s}{s}\frac{e^{-sz_{\beta,n}}}{\big(2\sin(s\lambda/2)\big)^{2-2r}}\;.
\end{eqnarray}
The topological string coupling $\lambda$ plays the role, in type IIA, of $g_s W^-$ where $g_s$ is the string coupling and $W^-$ the graviphoton background.  $\alpha^{{\cal \beta}}_r$ is the degeneracy of BPS states with central charge 
\be 
z_{\beta,n} = \beta\cdot t-2\pi in \;,
\ee 
and target space genus $r$. From the M-theory perspective, $n$ is the KK number of the M2 on the M-theory circle.

The integration over Schwinger proper time $s$ in (\ref{FP}) has an ultraviolet cutoff $\epsilon$. This can be taken arbitrarily small, though (\ref{FP}) is independent of the choice for $\epsilon$ within the range 
\be 
\label{Prangee}
0< \epsilon < 1 \;.  
\ee

Starting from the expression (\ref{FP}), we can perform the integral by Poisson resummation 
\be 
\sum_{n\in\mathbb{Z}}e^{2\pi i ns} = \sum_{k\in\mathbb{Z}}\delta(s-k) \;,
\ee
yielding the often-used form
\begin{eqnarray}
\label{sumk}
\label{Poisson}
   F^p= \sum_{k= 1}^{\infty}\sum_{{\bf \beta},r\geq 0}\alpha_r^{{\bf \beta}}\frac{e^{-k\beta\cdot t}}{k\big(2\sin(k\lambda/2)\big)^{2-2r}} \;.
\end{eqnarray}
Note that this expression should be understood as capturing only the perturbative asymptotic expansion in $\lambda$.

The integrating out calculation was performed in more detail in \cite{Dedushenko:2014nya} where it was shown that these BPS excitations of the M2-branes were the only ones that could possibly generate F-terms. This suggests that also the non-perturbative contributions to the free-energy should be captured by the integrating out calculation. In this work we show that this is indeed the case, but one must perform the integrating out calculation carefully. 

The main point is that the integrating out calculation does not yield (\ref{FP}), but in fact yields
\be
\label{Fnp}
    F = \sum_{{\bf \beta},r\geq 0,n\in \mathbb{Z}}\alpha_r^{{\bf \beta}}\int_{\epsilon}^{\infty}\frac{\text{d}s}{s}\frac{e^{-s|z_{\beta,n}|^2}}{\big(2\sin(\bar{z}_{\beta,n}s\lambda/2)\big)^{2-2r}} \;.
\ee
This is actually an intermediate expression in the calculation of \cite{Dedushenko:2014nya,Gopakumar:1998jq}. However, the steps to the final presented expression (\ref{FP}) are not valid in general, and so our proposal is that (\ref{Fnp}) is the formula which must be used. Further, we label (\ref{Fnp}) by $F$ rather than $F^p$ to show that it actually yields the full non-perturbative free energy. 

More precisely, the calculation yields the non-perturbative free energy in the so-called holomorphic limit \cite{Bershadsky:1993cx,Antoniadis:1993ze,Dedushenko:2014nya}. This neglects anti-holomorphic contributions to the free energy which on the supergravity side come from integrating out certain massless modes. These lead to an effective action which is not local and can violate the holomorphic properties of the Wilsonian action. Such non-holomorphic terms are not captured by the integrating out that we are performing. They are instead controlled by the so-called holomorphic anomaly equation \cite{Bershadsky:1993cx,Antoniadis:1993ze}. 

Related to this, the integrating out expression (\ref{Fnp}) holds in the case when the two-cycles wrapped by the M2 branes are rigid and smooth. We discuss this in more detail in section \ref{sec:evalf}. 

The formula (\ref{Fnp}) also has an ultraviolet cutoff $\epsilon$, but in our proposal this cutoff must be stronger than the range (\ref{Prangee}). We demand that $\epsilon$ must approach 0 from above
\be 
\epsilon = 0^+ \;.
\label{ep0p}
\ee 
We explain how to account for the $\epsilon =0$ value in section \ref{sec:0pole}, following the work presented in \cite{Hattab:2023moj}. 

While the integrands of (\ref{Fnp}) and (\ref{FP}) are related by a rescaling $s \rightarrow \bar{z}_{\beta,n} s$, they are not equivalent. This is because of the $n$ dependence in the rescaling, which means that after such a rescaling the integrating paths (in the complex plane) for each $n$ differ. Further, it is not possible to deform them all into each other because of the poles at $s = \frac{2\pi k}{\lambda}$ in (\ref{FP}) for any $k \in \mathbb{Z}$. These poles capture the non-perturbative physics.

\section{Evaluating the formula}
\label{sec:evalf}

Our proposal is that the full non-perturbative topological string free energy is given by the formula (\ref{Fnp}). In this section we explain how to evaluate it.  

The expression will only depend on the structure of the poles of (\ref{Fnp}) so we can focus on the case where the only non-zero invariant is $\alpha_{0}^{1} = 1$ and the general result will follow directly. With this restriction we can also, with no further restriction, consider only a single Kahler field $t$. We denote the real components explicitly as
\be
t = v + i b \;,
\ee
with $v > 0$.
This restriction corresponds to the case where Y is the resolved conifold, a non-compact Calabi-Yau which is a cone over a two-sphere and we denote its topological string free energy as $F^{rc}$.

To write $F^{rc}$ explicitly as a holomorphic function of $t$ we start from (\ref{Fnp}) and perform a change of variable 
\be 
u_n = s\bar{z}_{n} \;.
\ee
After the change of variable, the free energy is given by
\begin{eqnarray}
\label{angles}
    F^{rc} = \sum_{n\in\mathbb{Z}}\int_{0^+}^{\infty e^{i\theta_n}}\frac{\text{d}u}{u}\frac{e^{-uz_n}}{(2\sin\big(u\lambda/2)\big)^2}\;,
\end{eqnarray}
with $z_n = t-2\pi i n$, and 
\be 
\theta_n =\tan^{-1}\left(\frac{2\pi n-b}{v}\right) \;,
\ee 
is the argument of $\bar{z}_n$. 
Note that since $u_n$ is a dummy variable in the integral, we drop the $n$ subscript and just write $u_n \rightarrow u$.
By $\int_{0^+}^{\infty e^{i\theta_n}}$ we mean an integration along the line connecting $0^+$ to infinity at an angle $\theta_n$.

\begin{figure}
\centering
 \includegraphics[width=0.5\textwidth]{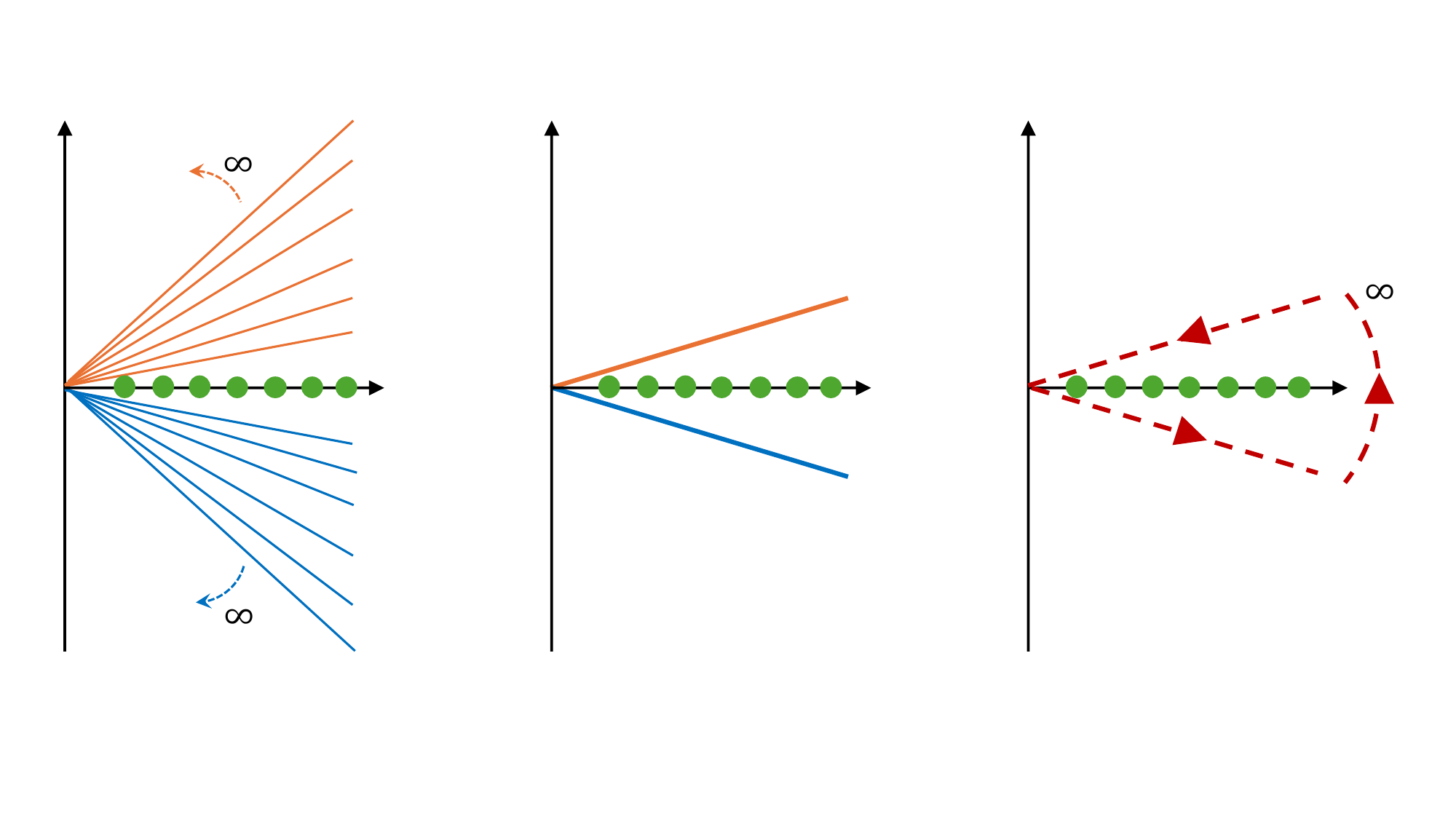}
\caption{Figure showing how the free energy (\ref{angles}) is a sum over different integration lines in the complex $u$ plane. For the choice $\lambda\in\mathbb{R}^+$ the poles are located on the real axis. It is possible to deform all the lines in the upper plane to coincide at some angle $\theta^+$, and the same for the lower plane with angle $\theta^-$. Finally, the integration path can be closed off at infinity yielding a contour integral which picks up all the poles.}
\label{fig:line}
\end{figure}

At this point it is useful to restrict $\lambda\in\mathbb{R}^+$. The evaluation for arbitrary $\lambda$ has certain subtleties that are not crucial for the discussion, and are discussed in more detail in section \ref{sec:stokes}. With this choice, the poles in (\ref{angles}) are all located on the real line in the $u$ plane. This is shown in figure \ref{fig:line} (in the first picture).

To perform the sum over $n$ we need the integrals to be on the same line. We can rotate the integration lines in the plane using contour integrals, noting that the integrand vanishes at infinity. However, we cannot rotate the lines through the real axis because of the poles. Therefore, the best we can do is to bring all integrals with $2\pi n-b>0$ onto the same line at angle $\theta_+\in (0,\pi/2)$ and similarly those with $2\pi n-b<0$ onto another line with angle $\theta_-\in (0,-\pi/2)$.
This is illustrated in figure \ref{fig:line} (in the second picture). After this rotation, we can write the sum over the integrals as
\begin{eqnarray}
    F^{rc} &=& \sum_{\{n\;:\;\theta_n>0\}}\int_{0^+}^{\infty e^{i\theta_+}}\frac{\text{d}u}{u}\frac{e^{-uz_n}}{\big(2\sin(u\lambda/2)\big)^2} \nonumber \\
    &+&\sum_{\{n\;:\;\theta_n<0\}}\int_{0^+}^{\infty e^{i\theta_-}}\frac{\text{d}u}{u}\frac{e^{-uz_n}}{\big(2\sin(u\lambda/2)\big)^2} \;.
\end{eqnarray}

To evaluate the sums over $n$, let us split $b$ as
\be
b = 2\pi n_b + \tilde{b} \;,
\ee
where $n_b = \floor{\frac{b}{2\pi}} \in \mathbb{Z}$ and $\tilde{b} \in (0,2\pi)$. Then the sums over the integrands give
\begin{eqnarray}
\sum_{\{n\;:\;\theta_n>0\}} e^{-u z_n} &=& -\frac{e^{-\left(t-2\pi i n_{b} \right)u}}{1-e^{-2\pi i u}} \label{sumnpos}\;, \\
\sum_{\{n\;:\;\theta_n<0\}} e^{-u z_n} &=& \frac{e^{-\left(t-2\pi i n_{b} \right)u}}{1-e^{-2\pi i u}} \;,
\label{sumnneg}
\end{eqnarray}
where we used that the imaginary part of $u$ has the right sign so the geometric sums are convergent. We then obtain
\begin{eqnarray}
     F^{rc} &=&-\int_{0^+}^{\infty e^{i\theta_+}}\frac{\text{d}u}{u}\frac{1}{1-e^{-2\pi i u}}\frac{e^{-\left(t-2\pi i n_{b} \right)u}}{\big(2\sin(u\lambda/2)\big)^2} \nonumber \\
    &+&\int_{0^+}^{\infty e^{i\theta_-}}\frac{\text{d}u}{u}\frac{1}{1-e^{-2\pi i u}}\frac{e^{-\left(t-2\pi i n_{b} \right)u}}{\big(2\sin(u\lambda/2)\big)^2} \;.
\end{eqnarray}

We can now combine these two integrals into a single closed contour integral 
\be
-\int_{0^+}^{\infty e^{i\theta_+}} + \int_{0^+}^{\infty e^{i\theta_-}} = \oint_C \;, 
\ee
where $C$ is a closed contour starting at $0^+$ and encircling all the poles on the real axis, in an anti-clockwise manner. We have used that the contribution to the integral from closing the contour at infinity vanishes. This is illustrated in figure \ref{fig:line} (in the third picture).

We then arrive at
\begin{eqnarray}
\label{rcres}
F^{rc}=\oint_C\frac{\text{d}u}{u}\frac{1}{1-e^{-2\pi i u}}\frac{e^{-\left(t-2\pi i n_{b} \right)u}}{\big(2\sin(u\lambda/2)\big)^2} \;.
\end{eqnarray}
Let us note that it was proposed already in \cite{Hattab:2023moj} that we should evaluate the prepotential through a contour integral in complex Schwinger proper time. We have now derived this. As shown in \cite{Hattab:2023moj}, this also helps to understand the zero pole, which was proposed to capture the ultraviolet completion of the sum over the Schwinger integrals. We discuss the zero pole in section \ref{sec:0pole}.

The integration over the contour $C$ in (\ref{rcres}) receives two types of contributions from poles. There are the poles coming from $u \in \mathbb{N}^*$. These give the perturbative (in $\lambda$) exponentials in the free energy (worldsheet instanton contributions to the prepotential). These are captured already by the original formula (\ref{FP}). The poles at 
\be 
u \in \left(\frac{2\pi}{\lambda} \right)\mathbb{N}^* \;,
\label{nppoles}
\ee 
are new and we propose that they calculate the non-perturbative contributions to the free energy. In the special case where $\lambda \in \pi \mathbb{Q}$ the two types of poles coincide at specific values, leading to third order poles.

Given the expression (\ref{rcres}), the general formula for an arbitrary Calabi-Yau now follows directly
\be
\label{full}
    F = \sum_{{\bf \beta},r\geq 0}\alpha_r^{{\bf \beta}}\oint_C\frac{\text{d}u}{u}\frac{1}{1-e^{-2\pi i u}}\frac{e^{-(\beta\cdot t-2\pi i n_{\beta\cdot b} )u}}{\big(2\sin(u\lambda/2)\big)^{2-2r}} \;,
\ee
where we now have $n_{\beta \cdot b} = \floor{\frac{\beta \cdot b}{2\pi}} \in \mathbb{Z}$.
We propose that (\ref{full}) gives the full non-perturbative topological string free energy (in the holomorphic limit, and for purely rigid two-cycles). This holds for real $\lambda$, and we discuss complex $\lambda$ in section \ref{sec:stokes}.

Note that the contribution from constant maps (pure D0 states), for any Calabi-Yau, is captured by the $t \rightarrow 0$ limit of (\ref{rcres}), times $-\frac{\chi(Y)}{2}$, where $\chi(Y)$ is the Euler characteristic of the Calabi-Yau. 

We can compare our result to the literature. The non-perturbative completion of the topological string partition function is a substantial topic of research. We refer to \cite{Gu:2023mgf,Alim:2024dyi,Alexandrov:2023wdj} for some of the most recent work, and \cite{Marino:2012zq,ANICETO20191} for reviews. 
In the specific case of the resolved conifold, the non-perturbative completion has been proposed using various means: by either studying matrix model dualities, Borel resummation techniques or solving difference equations \cite{Pasquetti:2010bps,Hatsuda:2013oxa,Krefl:2015vna,Hatsuda:2015owa,Alim:2021ukq}. It is given by

\begin{eqnarray}
\label{Flit}
    F^{rc} = F^p-\frac{\partial}{\partial \lambda}\left(\lambda F_{NS}\left[\frac{2\pi}{\lambda},\frac{2\pi}{\lambda}(t-i\pi)\right]\right) \;.
\end{eqnarray}
$F_{NS}$ corresponding to the NS limit \cite{Nekrasov:2009rc} of the refined topological string on the resolved conifold
\begin{eqnarray}
     F_{NS}\left[\lambda,t\right] = \frac{1}{2\pi}\sum_{k\geq 1}\frac{e^{-k t}}{k^22\sin(k\pi\lambda)} \;.
\end{eqnarray}
A simple evaluation of the integral (\ref{rcres}) for $\lambda\notin \pi\mathbb{Q}$ on the poles (\ref{nppoles}) gives, for $n_b=0$, precisely
\begin{eqnarray}
    -\frac{\partial}{\partial \lambda}\left(\lambda F_{NS}\left[\frac{2\pi}{\lambda},\frac{2\pi}{\lambda}(t-i\pi)\right]\right)\;.
\end{eqnarray}
We therefore find a match, at least for $n_b=0$.

We can also evaluate the integral (\ref{full}) over the poles (\ref{nppoles}), for $\lambda\notin \pi\mathbb{Q}$, yielding
\be
-\frac{\partial}{\partial \lambda}\left(\lambda \sum_{\beta}\alpha_0^{\beta}F_{NS}\left[\frac{2\pi}{\lambda},\frac{2\pi}{\lambda}(\beta\cdot t-2\pi in_{\beta\cdot b} -i\pi)\right]\right)\;.
\ee
This expression appears to match, for $b=0$, a conjecture recently put forward in \cite{Alim:2024dyi}. 

As stated, we restrict to the case where the two-cycles are rigid and smooth. This is related to the transformation between (\ref{rcres}) and (\ref{full}), specifically the $\alpha_r^{\beta}$. When the cycles are rigid and smooth, this transformation is a simple multiplicity counting. However, if the cycles are not rigid or smooth, the transformation is associated to tracing over additional fermionic zero modes \cite{Gopakumar:1998jq,Dedushenko:2014nya}. It can be, and we expect so, that this trace is more refined at the non-perturbative level. This will be discussed in \cite{nonrigid}. We expect that accounting for this is needed to match results in the literature, such as in \cite{Hatsuda:2013oxa}, which do not match (\ref{full}) and are associated to non-rigid geometries. We note, however, that for any Calabi-Yau geometry, the expression (\ref{full}) captures the rigid contributions. Also, we expect it to capture the leading non-perturbative terms in the $\lambda \rightarrow \infty$ limit, as studied in \cite{Hattab:2024yol}. 

\section{Pole crossing jumps}
\label{sec:stokes}

The expression (\ref{Fnp}) also captures observations from resurgence considerations. Namely that the non-perturbative terms can be identified with Stokes jumps from a Borel summation \cite{Pasquetti:2010bps,Hatsuda:2015oaa,Alim:2024dyi,Alim:2021mhp}. To show how this features in expression (\ref{Fnp}), we consider the case where $\lambda$ is complex, $\lambda = \lambda_r+i\lambda_i$. For simplicity, we restrict $\lambda_r\geq 0$. This has the effect of rotating the line with the non-perturbative poles (\ref{nppoles}) around the origin in the right half plane. We denote the angle of this poles half-line $\theta_{\lambda}$ and the half-line itself $l_{\lambda}$ (we exclude the origin from this line). We also denote the integration half-lines with angles $\theta_n$ as $l_{n}$.

Like in section \ref{sec:evalf}, we focus on the resolved conifold, since the general case follows straightforwardly.
If $\theta_{n_b+1}>\theta_{\lambda}>\theta_{n_b}$ then the result is left unchanged and is still given by (\ref{rcres}). However, if the poles on $l_{\lambda}$ cross one of the $l_{n}$, then the analysis differs. In order to deform all the $l_n$ with $n>n_b$ onto the same line (or $n\leq n_b$ depending on the sign of $\lambda_i$), we will need to have some $l_n$'s crossing $l_{\lambda}$. Each such line which crosses $l_{\lambda}$ will contribute one contour integral around the poles of $l_{\lambda}$.

Say that for some $k \in \mathbb{N}^*$ we have $\theta_{n_b+k+1}>\theta_{\lambda}>\theta_{n_b+k}$. Then in order to perform the sums as before, we have to bring all the lines $l_{n_b+1}$ to $l_{n_b+k}$ into the angle $\theta_+\in(\theta_\lambda,\pi/2)$. We do this by including for each line the extra contour over the poles (\ref{nppoles}) on the line $l_{\lambda}$
\begin{eqnarray}
    F^{rc}&=&\oint_{C}\frac{\text{d}u}{u}\frac{1}{1-e^{-2\pi i u}}\frac{e^{-(t-2\pi in_b)u}}{\big(2\sin(u\lambda/2)\big)^2} \nonumber\\
    &+& \sum_{j=1}^k \oint_{C_{\lambda}}\frac{du}{u}\frac{e^{-\left(t-2\pi in_b-2\pi i j\right)u}}{\big(2\sin(u\lambda/2)\big)^2}\;,
    \label{changelin}
\end{eqnarray}
where the contour $C$ now surrounds both the poles on the real axis and the poles on $l_{\lambda}$ (not including the zero pole). The contour $C_{\lambda}$ only surrounds the poles on $l_{\lambda}$.
Evaluating one of the integrals inside the sum over $j$ in (\ref{changelin}) gives  
\be
\frac{\partial}{\partial \lambda}\bigg(\frac{\lambda}{2\pi i}\text{Li}_2\left(e^{-\frac{2\pi}{\lambda}\left( t-2\pi in_b-2\pi i j\right)}\right)\bigg)\;.
\ee
These are known as Stokes jumps.

In the extreme case where we send $k$ to infinity, which is equivalent to taking the topological string coupling to be purely imaginary $\lambda\in i\mathbb{R}^-$, the sum over $j$ is now exactly the same as the sum in (\ref{sumnpos}). This means that the non-perturbative poles from the contour $C$ will cancel with the sum over the Stokes jumps completely. The answer we end up with is then just the perturbative piece $F^p$ that can be computed from (\ref{FP}). From the perspective of the line integrals, this is expected since having $\lambda$ pure imaginary means we can bring all the integration rays onto the real line without crossing any poles.

The general formula (\ref{Fnp}) features exactly the same jumps
\be
\alpha^{\beta}_0\frac{\partial}{\partial \lambda}\bigg(\frac{\lambda}{2\pi i}\text{Li}_2\left(e^{-\frac{2\pi}{\lambda}\left(\beta\cdot t -2\pi i n_{\beta\cdot b} - 2 \pi i j\right)}\right)\bigg)\;.
\ee
To evaluate the general formula (\ref{Fnp}), for complex $\lambda$, one needs to account for these Stokes jumps, and after that it is possible to use (\ref{full}). 

While complex $\lambda$ manifests some interesting properties, physically it is not clear what the integrating out calculation means for complex $\lambda$, and so it is not clear that these properties are really physically correct. We leave a better understanding of complex $\lambda$ physics for future work.

\section{The zero pole}
\label{sec:0pole}

Throughout this work we have excluded the zero pole by setting an ultraviolet cutoff in the Schwinger proper time parameter (\ref{ep0p}). The cutoff is arbitrarily close to the zero pole, but does not include it. It was already proposed in \cite{Hattab:2023moj,Hattab:2024thi} that the zero pole captures the ultraviolet physics of the integrating out procedure, and further, that it induces the polynomial terms in the free energy. These polynomial terms are missing from the expression (\ref{Fnp}), because the $\epsilon$ cutoff removes the zero pole. 

It is straightforward to include the zero pole for the case of the resolved conifold, as was already done in \cite{Hattab:2023moj}. The pole yields polynomial contributions to $F_0$ and $F_1$ given (for $n_b=0$) by
\begin{eqnarray}
F^{\mathrm{Poly}}_0 &=& -\frac{1}{6}t^3 + \frac{\pi i}{2}t^2 + \frac{\pi^2}{3}t \;,	\\
F^{\mathrm{Poly}}_1 &=& -\frac{1}{12}t + \frac{\pi i}{12} \;.
\end{eqnarray}
These match the expected results coming from a geometric transition from Chern-Simons theory \cite{Gopakumar:1998ki}, up to a factor of $\frac12$.\footnote{The constant piece in $F_1$ is cancelled by the pure D0 contribution from the same zero pole, and is absent on the Chern-Simons side.} The factor of $\frac12$ was discussed in \cite{Hattab:2023moj}, and further in \cite{Hattab:2024chf}. It arises because for the resolved conifold the zero pole has a degeneracy of $2$ associated to capturing both sides of the flop transition. The negative definite poles in the resolved conifold case should be associated to the flopped manifold, and the zero pole is shared by both sides. In compact Calabi-Yau manifolds, such a degeneracy does not occur around the large volume point. 

The fact that the zero pole gives the tree-level polynomial terms is presented in \cite{Hattab:2023moj} as evidence for the Emergence Proposal \cite{Palti:2019pca,Grimm:2018ohb,Heidenreich:2017sim,Heidenreich:2018kpg}. However, as explained in \cite{Hattab:2023moj,Hattab:2024thi,Hattab:2024chf}, the emergence is in general more subtle because the ultraviolet is in a different phase to the infrared. This means that in general, so for cases other than the resolved conifold, it is not true that the ultraviolet contribution can be evaluated directly from the zero pole of (\ref{Fnp}). Indeed, in the ultraviolet phase one cannot treat the different wrapped branes as independent particles, but rather one must consider a single set of constituent degrees of freedom which capture all the tower of wrapped branes (and more). Such a description is known for some special cases, such as the blow-up conifold, and is discussed in \cite{Hattab:2024thi,Marino:2011eh}. As proposed in \cite{Hattab:2023moj,Hattab:2024thi}, for compact cases it is still possible to associate the leading polynomial contributions to a zero pole, but this pole is in the Mellin-Barnes representation of the period, rather than the prepotential. 

Nonetheless, we expect that the contribution from the zero pole which is omitted in (\ref{Fnp})  will always correspond to the polynomial terms in $F_0$ and $F_1$. The full, non-perturbative, topological string free energy is therefore given by (\ref{Fnp}), supplemented by polynomial terms (as well as the pure $D0$ contribution proportional to the Euler number of the Calabi-Yau). These can be deduced from direct dimensional reduction in the compact Calabi-Yau cases, and from geometric transitions in the non-compact cases.

\vspace{0.1cm}
{\bf Acknowledgements}
\noindent
 The work of JH and EP is supported by the Israel Science Foundation (grant No. 741/20) and by the German Research Foundation through a German-Israeli Project Cooperation (DIP) grant ``Holography and the Swampland". The work of EP is supported by the Israel planning and budgeting committee grant for supporting theoretical high energy physics.

\bibliography{Higuchi}

\end{document}